# Refrain from adopting the combination of citation and journal metrics to grade publications, as used in the Italian national research assessment exercise (VQR 2011-2014)[1]

*Giovanni Abramo* (corresponding author)
Laboratory for Studies of Research and Technology Transfer
Institute for System Analysis and Computer Science (IASI-CNR)
National Research Council of Italy
Via dei Taurini 19, 00185 Rome - ITALY
    giovanni.abramo@uniroma2.it
    Tel/fax +39 06 72597362

*Ciriaco Andrea D'Angelo*
Department of Engineering and Management
University of Rome "Tor Vergata"
Via del Politecnico 1, 00133 Rome - ITALY
    dangelo@dii.uniroma2.it

**Abstract**

The prediction of the long-term impact of a scientific article is challenging task, addressed by the bibliometrician through resorting to a proxy whose reliability increases with the breadth of the citation window. In the national research assessment exercises using metrics the citation window is necessarily short, but in some cases is sufficient to advise the use of simple citations. For the Italian VQR 2011-2014, the choice was instead made to adopt a linear weighted combination of citations and journal metric percentiles, with weights differentiated by discipline and year. Given the strategic importance of the exercise, whose results inform the allocation of a significant share of resources for the national academic system, we examined whether the predictive power of the proposed indicator is stronger than the simple citation count. The results show the opposite, for all discipline in the sciences and a citation window above two years.

# 1. Introduction

In the current knowledge economy, a country's research infrastructure plays an increasingly strategic role in supporting competitiveness and socioeconomic development. A growing number of countries carry out periodic national research assessment exercises to monitor their research systems, stimulate continuous improvement, and assess the impact of policy initiatives adopted to improve them. The purpose of the research assessments is to support all or some of the following goals: to inform research policies at national and regional levels; to inform strategic planning at the institutions level; to selectively allocate funding; to stimulate continuous improvement at individual and organization levels; to reduce the information asymmetry between suppliers of knowledge (research institutions) and demand (students, companies); and last but not least, to demonstrate that investment in research is effective and delivers public benefits. Given such important goals, one would expect the assessment exercises to be formulated and executed in keeping with the state of the art in research evaluation. However, various studies have demonstrated that the truth is at times far different, and this is certainly the case for Italy (Abramo, D'Angelo, and Caprasecca, 2009; Abramo, Cicero, and D'Angelo, C.A., 2013; Abramo, D'Angelo, and Di Costa, 2014; Abramo and D'Angelo, 2015; Baccini and De Nicolao, 2016; Baccini, 2016). Indeed the cure, when based on the results of such assessments, can be far worse than the ailment (Butler, 2003a and 2003b; Abramo, D'Angelo, and Di Costa, 2011; Abramo, Cicero, and D'Angelo, C.A., 2011).

After the first Italian national research assessment exercise (VTR 2001-2003) run by an ad hoc committee (CIVR), the government decided to appoint a permanent body, the Italian Agency for the Evaluation of Universities and Research Institutes (ANVUR), with the mission to run periodic national research assessment exercises. ANVUR started its operations in 2011, launching the second national research assessment exercise (VQR 2004-2010). The VQR methodology began soon the object of a wide debate in the Italian academic community in blogs (www.roars.it; www.lavoce.info), newspapers and magazines. Notwithstanding the strong criticisms raised in the scientific arena to the methods and indicators used in the VQR 2004-2010 (Franco, 2013; Baccini and De Nicolao, 2016; Baccini, 2016; Abramo, D'Angelo, and Di Costa, 2014; Abramo and D'Angelo, 2015), the third exercise (VQR 2011-2014, with results expected by the end of 2016) adheres to exactly the same framework, in almost total disregard of both the criticisms and recommendations for improvement. The only difference that can be noted concerns the bibliometric indicator used for scoring the publications submitted by the research institutions. Both in the past and current editions of the VQR, the quality score for each publication derives from a combination of citations (C) and journal (J) metrics. In the previous VQR, the (C,J) space was partitioned into quality blocks by using discrete thresholds for both citations and journal metrics separately. In the current one, the thresholds are determined through a linear weighted combination of the C world percentile rank (for publications of the same year and subject category) and the J percentile rank.

The new C-J combined indicator methodology, used in the VQR 2011-2014[2], was conceived by an ANVUR research team and presented in this same journal (Anfossi et al., 2016)[3]. The authors depicted it as "intuitive and versatile; effective in terms of cost and time; and the derived evaluation as homogeneous among different scientific domains".

---

[2] http://www.anvur.org/attachments/article/825/Bando%20VQR%202011-2014_secon~.pdf (last accessed 23/09/2016).

We object, stating that in fact the combined indicator is not as able at predicting the future impact of publications as the much simpler, widely accepted citation count indicator. Moreover, the C-J indicator appears to have been formulated without rigorous scientific method, and instead as a flight of fancy. Finally, contrary to what the proponents affirmed, the implementation resulted as more cumbersome, costly and time consuming than that of alternate methodologies, particularly for the administrations and researchers tasked with selecting their best two or three products according to such an indicator. For the benefit of those responsible for future evaluations, and for the achievement of national, institutional and individual academic goals, we demonstrate that the simple citation count is a better proxy of a publication impact than the C-J metric adopted by ANVUR.

## 2. The combination of citations and journal metrics applied in the Italian VQR (2011-2014)

In the paper proposing the C-J combined indicator, Anfossi et al. (2016) observed, with good reason, that "the use of the sole citation count may not be an appropriate indicator of impact in those cases where the paper is too young". They then propose a combination of journal metrics and article citations as a "potentially powerful tool for compensating intrinsic flaws of citation count alone".

Our first question is when is a paper "too young" for the use of citation count to predict its future impact, and thus potentially better assessed by the combination of C and J metrics. In fact in the literature, the hybrid indicator has only been recommended for citation windows of zero or one years (Levitt & Thelwall, 2011), although its use for a two-year window has been suggested for papers in mathematics (and with weaker justification in biology and earth sciences), because of the characteristic inertia of this discipline regarding the early stages of accruing citations (Abramo, D'Angelo & Di Costa, 2010). Confirming Levitt and Thelwall (2011), in the social sciences, the journal's impact factor is seen to improve the correlation between predicted and actual ranks by citation when applied in the "zero" year of publication and up to one year afterwards (Stern, 2014).

The most recent VQR referred to the period 2011-2014, with citations to be counted at the end of February 2016. Thus, the literature would recommend that the impact of all publications falling in the first three years of the period observation be measured by citations only. However, in the paper advancing the C-J indicator (Anfossi et al. 2016), there is no mention or apparent consideration of the above references. Still, any new form of measure that could better predict future impact, including a hybrid indicator, would be welcome by the bibliometrics community. Indeed, Anfossi et al. propose a combination of both C and J ranking metrics, with the weighting varying by discipline and year. As noted, this proposal was not justified on the basis of the literature. Still worse, there appears to be no attempt to develop empirical demonstration of this combined indicator as a valid predictor of impact, which might thus provide grounds for such recommendation. The authors state: "One can develop an algorithm which combines those parameters automatically, … or can find a way to directly choose the weight. Our view, … is that this degree of freedom should be left to the panel of experts because of different habits of scientific communities and because of different significance of citation count when applied to recent papers".

---

[3] At time of publication, all co-authors but Giorgio Parisi were affiliated to ANVUR. Giorgio Parisi was president of the panel of experts in physics in the VQR 2004-2010.

We object that there is no degree of freedom when it comes to quantitatively assessing the impact of publications. Simply put, there are indicators that work better than others, and given that such proofs are available, only rigorous scientific method should be used to choose among the good and bad options. Applying such scientific method, we will prove that the C-J combined indicator presented by Anfossi et al. (2016) and used in the VQR 2011-2014, provides a worse prediction of the impact of publications than the simple citation count.

The citation from Anfossi et al. (above) mentions the panels of experts set up to assess the research products submitted to the VQR (2011-2014). There were 16 such panels, named GEVs[4], reflecting the 14 University Disciplinary Areas (UDAs) in which Italian professors are classified. Two UDAs (Numbers 8, 11) were each split into two and the corresponding GEVs (for a total of four) were set up, thus allowing the use of different evaluation methods for research products from distinct areas. In four GEVs in the social sciences, arts and humanities and law,[5] the panels applied peer review as the exclusive method of evaluation. A further GEV (No. 13, Economics and statistics) chose to evaluate the articles based on the prestige of the publishing journals. Thus there remained 11 GEVs that applied bibliometrics to inform their assessment. Each publication submitted to evaluation in these GEVs was positioned in a C-J[6] space, such as the one shown in Figure 1. The C-IF space was partitioned in five regions by drawing four threshold curves. The *n-th* threshold is identified by setting a generic function equal to zero:

$$f_n(J,C) = Const_n + \frac{a}{1-a} J + \frac{1}{1-a} C$$

[1]

where:
a = slope of the curve;
C = publication's citation percentile, obtained by ranking in decreasing order the total number of publications indexed in WoS (or Scopus) of the same year and subject category (SC);
J = journal metric percentile, obtained by ranking in decreasing order by journal metric the journals belonging to the same SC as the journal publishing the publication.

The threshold curves are such that region A includes only publications ranking in the world top 10% by the C-J combined indicator (score = 1); region B includes those in the top 70% to 90% (score = 0.7); C between 50% and 70% (score = 0.4), D 20%-50%

---

[4] *Gruppo di Esperti della Valutazione* (Group of Experts in Evaluation).
[5] Specifically GEV 10 - Ancient history, philology, literature and art; GEV 11a - History, philosophy, pedagogy; GEV 12 - Law; and GEV 14 - Political and social sciences.
[6] The journal metrics proposed were: the 2- and 5-year Impact Factor (5YIF) and the Article Influence score (AI), for WoS indexed publications; the Impact per Publication (IPP) and the SCImago Journal Rank (SJR), for Scopus indexed publications.

(score = 0.1), and E the bottom 20% (score = 0).[7] Regions A to E were further labeled "Excellent" to "Very Poor" (Figure 1).

*Figure 1: The combination of citation and journal metric percentiles adopted by ANVUR to rate publications in the VQR (2011-2014). A=Excellent; B=Good; C=Adequate; D=Poor; E=Very poor*

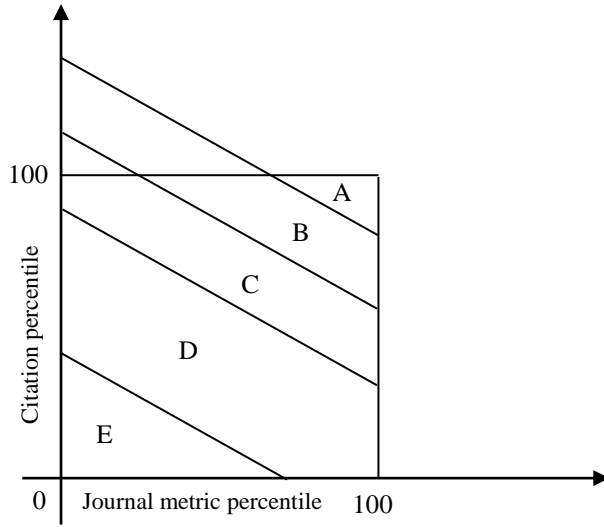

Each GEV decided the intercepts and the slope of the curves (the value of *a*), i.e. the weight assigned to C and IF to predict the impact of the publications. Table 1 shows the chosen values of slope, per GEV and year of publication (2011-2013). Note that all 2014 publications were evaluated only by peer review, except those falling in the top world 10% ("A" region, in Figure 1).

*Table 1: Slope values (a) of the threshold curves, by year, as chosen by GEVs*

| GEV | 2011 | 2012 | 2013 |
|---|---|---|---|
| 1 - Mathematics and computer science | -1.1 | -1.4 | -1.7 |
| 2 - Physics | -0.4 | -0.6 | -0.9 |
| 3 - Chemistry | -0.4 | -0.6 | -0.8 |
| 4 - Earth sciences | -0.4 | -0.6 | -0.9 |
| 5 - Biology | -0.4 | -0.6 | -0.8 |
| 6 - Medicine | -0.4 | -0.6 | -0.8 |
| 7 - Agricultural and veterinary sciences | -0.7 | -0.9 | -1.5 |
| 8a - Civil engineering | -0.6 | -0.9 | -1.5 |
| 8b - Architecture | -0.7 | -0.9 | -1.5 |
| 9 - Industrial and information engineering | -0.4 | -0.6 | -0.9 |
| 11b - Psychology | -0.4 | -0.6 | -1.0 |

---

[7] Note that the C-J space ultimately adopted under the 2011-2014 VQR differs in two aspects from the one presented by Anfossi et al. (2016). The procedure applied five regions instead of four. Further, each GEV was requested to distinguish two separate "peer review" regions, one at the top-left corner (high citation percentile/low journal metric percentile) and the other at the bottom-right corner (low citation/ high journal metric). The publications falling in these special regions were to be evaluated by informed peer-review.

## 3. Data and methods

We proceed as follows to examine whether the C-J combined indicator applied under ANVUR predicts the long-term impact of publications better than citation counts alone. The goal is to compare the results of C-J evaluation to those achieved by citation counts in terms of predictive power. To do this we need a reliable benchmark, which would be an evaluation based on citations, but counted several years after the date of publication. Operationally, we translate the period of observation for the VQR to a previous period. The VQR evaluated the publications over the years 2011-2014 for each professor on university staff at 31/12/2015, measuring the citations as accrued by early 2016. For our analysis we instead evaluate the publications of the period 2004-2006[8] for the professors on staff at the end of 2008. The citations are counted as of 31/12/2015 for the benchmark measurement, and as of 31/12/2008 in the other cases. For reasons of robustness, the analysis excludes two of the eleven GEVs that applied the C-J evaluation metric.[9] The field of observation thus consists of 30,595 professors who authored over 79,000 WoS-indexed publications, distributed per GEV as presented in Table 2.

*Table 2: Dataset for the analysis - Italian professors and their authorship of WoS 2004-2006 publications by GEV*

| GEV | Professors | Authorships | Publications |
|---|---|---|---|
| 1 - Mathematics and computer science | 2,868 | 7,616 | 5,672 |
| 2 - Physics | 2,183 | 30,648 | 11,769 |
| 3 - Chemistry | 2,715 | 26,321 | 13,326 |
| 4 - Earth sciences | 1,056 | 3,043 | 2,120 |
| 5 - Biology | 4,307 | 25,072 | 14,504 |
| 6 - Medicine | 9,431 | 55,225 | 26,455 |
| 7 - Agricultural and veterinary sciences | 2,672 | 8,106 | 4,326 |
| 8a - Civil engineering | 1,319 | 2,136 | 1,534 |
| 9 - Industrial and information engineering | 4,044 | 15,593 | 10,046 |
| Total | 30,595 | 173,760 | 79,273† |

† *Total is less than the sum of the column data due to multiple counts of publications co-authored by academics pertaining to more than one GEV. Co-authors belonging to different universities could in fact submit the same publication.*

For each publication we measure the values of three indicators of impact: the benchmark, the one applied under ANVUR, and that of citations only applying the ANVUR citation window:
1) $C_{long}$ = percentile (0 worst; 100 best) calculated on the basis of the comparison between the citations received by the publications as of 31/12/2015 and the citations received by all publications at the world level for the same year and subject category;[10]
2) $C_{short}$ = as above, but with citations counted as of 31/12/2008;
3) $\text{C-J} = \frac{a}{1-a} \text{IF} + \frac{1}{1-a} C_{short}$

---

[8] As stated above, most of 2014 publications are not evaluated by metrics.
[9] We have excluded GEV 8b (Architecture) and GEV 11b (Psychology) due to the limited share of products indexed in bibliometric databases out of the total research production of these professors in the period 2004-2006.
[10] For publications in multi-category journals the percentile considered is the one referring to the most favorable subject category.

Where:
- $IF$ = $IF$ percentile of the journal in which the article was published (evaluated for the year of publication), obtained by comparison of the impact factor of the journal of the publication and that of all the journals of the same subject category for the same year;
- a = slope of the threshold curves defined by GEV and year.

### 4. Analysis and discussion

In this section, we verify which of the two indicators, $C_{short}$ or C-J, offers better prediction of the long-term impact of publications. We conduct the verification in two steps. We first conduct the analysis comparing the grading by $C_{short}$ and C-J with that by $C_{long}$, for all the publications produced by the professors of the dataset. We then restrict the evaluation to the best two publications per professor, as provided for under the VQR 2011-2014.

#### 4.1 Analysis of the total publications in the dataset

As an example, Figure 2 provides a comparison of the three indicators of impact for the total of 40 publications in 2004-2006 by a full Professor of Experimental physics at the University of Ferrara (John Doe). The primary Y axis (at left) refers to the values of $C_{long}$ for these publications, while the secondary axis (at right) shows, respectively for each publication, the difference of C-J and $C_{short}$ from $C_{long}$. We note that both of these indicators are well able to approximate the impact of the publications measured over the long term, for those publications where $C_{long}$ is not less than 10. However for the publications that are little or not at all cited (i.e. $C_{long}$ is less than 10) the shifts are very substantial and similar for both $C_{short}$ and C-J.

*Figure 2: Impact values for all publications of John Doe, full professor of Experimental physics*

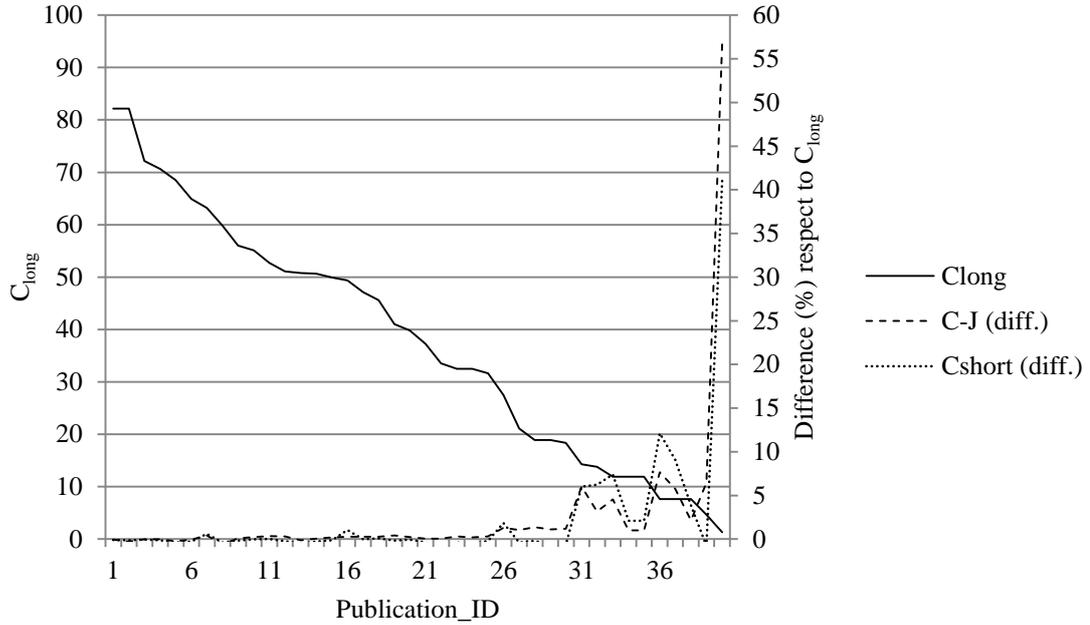

Figure 3 provides a further example of comparison between the three impact indicators, in this case for the total 46 publications of Jane Doe, Associate professor of Internal medicine at the University of Naples 'Federico II'. Here, we note the error committed in using C-J to estimate the impact measured over the long period ($C_{long}$) would be systematically higher than the error when using $C_{short}$. Moreover, while for $C_{short}$ the error is almost never greater than 10% (occurring for only one publication), using C-J the error is very substantial, above all for publications receiving scarce citations over the long period (i.e. with $C_{long}$ less than 10).

*Figure 3: Impact values for all publications of Jane Doe, associate professor of Internal medicine*

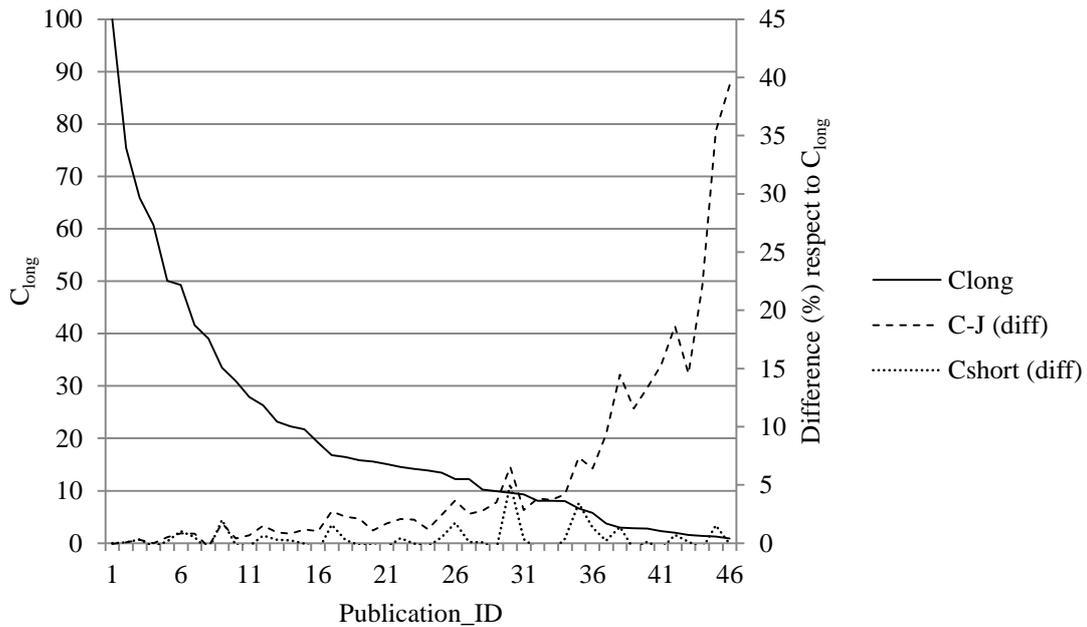

The dataset does not contain any examples of cases opposite to that of Jane, meaning a publications portfolio where C-J provides a better prediction of $C_{long}$. In effect, aggregating all the observations (89,752),[11] we obtain the correlations shown in Table 3, which indicate a level of convergence between $C_{long}$ and $C_{short}$ (Pearson and Spearman correlations equal to 0.497 and 0.484) greater than that between $C_{long}$ and C-J (0.377 and 0.400). The average value of the percentile differences between $C_{long}$ and C-J is 24.9; between $C_{long}$ and $C_{short}$ it is 22.4. The median is respectively at 21.6 and 13.8; while the differences dispersion (measured through standard deviation) is greater between $C_{long}$ and $C_{short}$ (24.0) than it is between $C_{long}$ and C-J (18.4).

*Table 3: Correlations and descriptive statistics of the distributions of percentile differences between C-J and $C_{long}$, and between $C_{short}$ and $C_{long}$*

|  | C-J vs $C_{long}$ | $C_{short}$ vs $C_{long}$ |
|---|---|---|
| Pearson correlation | 0.377 | 0.497 |
| Spearman correlation | 0.400 | 0.484 |
| Average difference | 24.9 | 22.4 |
| Median | 21.6 | 13.8 |
| Mode | 10-20 (22.8%) | 0-10 (38.7%) |
| Standard deviation | 18.4 | 24.0 |
| Curtosis | 1.014 | 1.822 |
| Skewness | 1.045 | 1.561 |
| Min-Max | 0-100 | 0-100 |

Subdividing the dataset of publications by year (Table 4), we note (as we would expect) that the levels of convergence of the annual distributions decline with the reduction of the citation window. However, $C_{short}$ always approximates $C_{long}$ better than C-J. In the comparison between $C_{short}$ and $C_{long}$, the Pearson correlation coefficient shifts from 0.543 in 2004 to 0.458 in 2006, while in the comparison between C-J and $C_{long}$ it drops from 0.481 a 0.273. Also, the average (and the median) of the differences (in absolute value) of the values of scores always favor $C_{short}$ over C-J; moreover, these differences increase as the citation window is shortened. It should be noted that the mode of the percentile difference distributions, always between 0-10, has much higher frequencies in the comparison between $C_{short}$ and $C_{long}$. Finally, the predictive power of C-J seems extremely low for 2006 publications (Pearson $\rho = 0.273$) indicating that, in the empirical framework adopted, the weight given to IF vis-à-vis citations was clearly too high.

The analysis by GEV (Table 5) confirms the stronger predictive power of $C_{short}$ vs C-J in every GEV, as shown by the Pearson indexes although, because of few outliers, in Industrial engineering and Mathematics and computer sciences, the average $C_{long}$ vs $C_{short}$ percentile difference is higher that the average $C_{long}$ vs C-J. In six out of nine GEVs the Pearson correlation between C-J and $C_{long}$ is below 0.4. The same occurs for $C_{long}$ vs $C_{short}$ in only four GEVs.

---

[11] For this analysis we eliminate the double counts of publications co-authored by professors pertaining to the same GEV. We consider instead publications co-authored by professors of different GEVs, because each GEV adopts different thresholds.

*Table 4: Correlations and descriptive statistics of the distributions of percentile differences between C-J and $C_{long}$, and between $C_{short}$, and $C_{long}$ per year*

|  | 2004 | | 2005 | | 2006 | |
| --- | --- | --- | --- | --- | --- | --- |
|  | C-J vs $C_{long}$ | $C_{short}$ vs $C_{long}$ | C-J vs $C_{long}$ | $C_{short}$ vs $C_{long}$ | C-J vs $C_{long}$ | $C_{short}$ vs $C_{long}$ |
| Obs. | 29,475 | | 29,655 | | 30,622 | |
| Pearson correlation | 0.481 | 0.543 | 0.398 | 0.513 | 0.273 | 0.458 |
| Spearman correlation | 0.496 | 0.522 | 0.426 | 0.500 | 0.296 | 0.456 |
| Average difference | 21.3 | 18.6 | 24.4 | 21.1 | 28.7 | 27.4 |
| Median | 17.8 | 10.5 | 22.1 | 13.1 | 26.5 | 19.4 |
| Mode | 0-10 (28.5%) | 0-10 (48.5%) | 0-10 (23.4%) | 0-10 (41.6%) | 0-10 (19.6%) | 0-10 (30.4%) |
| Standard deviation | 17.5 | 22.6 | 17.3 | 23.1 | 19.5 | 25.3 |
| Curtosis | 2.753 | 3.716 | 1.022 | 2.566 | 0.199 | 0.463 |
| Skewness | 1.546 | 2.022 | 0.968 | 1.729 | 0.724 | 1.130 |
| Min-Max | 0-100 | 0-100 | 0-100 | 0-100 | 0-99.3 | 0-100 |

*Table 5: Correlations and descriptive statistics of the distributions of percentile differences between C-J and $C_{long}$, and between $C_{short}$ and $C_{long}$ per GEV*

|  |  | Pearson correlation | | Average difference | | Standard deviation | |
| --- | --- | --- | --- | --- | --- | --- | --- |
| GEV | Obs | C-J vs $C_{long}$ | $C_{short}$ vs $C_{long}$ | C-J vs $C_{long}$ | $C_{short}$ vs $C_{long}$ | C-J vs $C_{long}$ | $C_{short}$ vs $C_{long}$ |
| 1 | 5,672 | 0.077 | 0.347 | 30.5 | 31.3 | 20.3 | 28.9 |
| 2 | 11,769 | 0.320 | 0.347 | 27.7 | 25.5 | 20.8 | 27.6 |
| 3 | 13,326 | 0.477 | 0.568 | 23.1 | 19.4 | 16.6 | 20.8 |
| 4 | 2,120 | 0.310 | 0.439 | 23.1 | 19.4 | 19.0 | 24.0 |
| 5 | 14,504 | 0.507 | 0.628 | 21.8 | 18.0 | 16.3 | 19.8 |
| 6 | 26,455 | 0.502 | 0.632 | 21.9 | 18.3 | 16.2 | 20.1 |
| 7 | 4,326 | 0.215 | 0.516 | 28.8 | 24.2 | 18.3 | 24.3 |
| 8a | 1,534 | 0.176 | 0.383 | 30.8 | 29.9 | 20.0 | 26.3 |
| 9 | 10,046 | 0.199 | 0.250 | 30.0 | 32.8 | 21.6 | 29.0 |

*Legend: GEV 1 - Mathematics and computer science; 2 – Physics; 3 – Chemistry; 4 - Earth sciences; 5 – Biology; 6 – Medicine; 7 - Agricultural and veterinary sciences; 8a - Civil engineering; 9 - Industrial and information engineering*

## 4.2 A comparison of the predictive power of C-J and $C_{short}$ at identifying the best two publications per professor

For reasons of cost and time, peer-review evaluation exercises generally provided for the selection and consideration of only the "best" publications. In the Italian VQR, each institution under evaluation (universities, research institutions under the Ministry of University Education and Research) was requested to select and submit the two best research products achieved in the 2011-2014 period, for each professor. An institution or professor attempting an effective selection process would obviously refer to the A-E grading seen in Figure 1. Applying this categorization to the 173,760 authorships of the dataset on the basis of their relative percentiles under $C_{short}$ and $C_{long}$, we obtain the results described in Table 6.

The cells along the main diagonal represent the "correct" classification, meaning where the grading obtained using $C_{short}$ coincides with that obtained using $C_{long}$. The cells above the diagonal represent the cases of "overgrading" (where $C_{short}$ would give a better grading than that from $C_{long}$). Finally, the cells below the main diagonal represent the cases of "undergrading" (where $C_{short}$ would give a worse grading than that from $C_{long}$). The correctly graded cases are 44.2% of the total, those that are overgraded represent 36.2 % and those that are undergraded represent 19.6% of the total. Repeating this analysis with C-J (Table 7), the correctly graded cases drop to 28.4% of total, above all because of the frequency of cases of overgrading (54.0%).

*Table 6: Evaluation of the 173,760 authorships of the dataset according to the VQR grading (A=excellent; B= good; C=adequate; D=poor E=very poor), but on the basis of $C_{short}$ and $C_{long}$*

|  | | $C_{long}$ | | | | |
|---|---|---|---|---|---|---|
| | Grade | A | B | C | D | E |
| $C_{short}$ | A | 7,531 | 10,065 | 7,515 | 8,231 | 6,049 |
| | B | 1,444 | 6,718 | 4,999 | 3,291 | 891 |
| | C | 347 | 5,226 | 9,223 | 10,492 | 1,831 |
| | D | 319 | 3,506 | 9,882 | 27,444 | 9,454 |
| | E | 1,351 | 1,774 | 1,257 | 8,971 | 25,949 |

*Table 7: Evaluation of the 173,760 authorships of the dataset according to the VQR grading (A=excellent; B= good; C=adequate; D=poor E=very poor), but on the basis of C-J and $C_{long}$*

|  | | $C_{long}$ | | | | |
|---|---|---|---|---|---|---|
| | Grade | A | B | C | D | E |
| C-J | A | 1,661 | 2,814 | 2,679 | 3,112 | 1,730 |
| | B | 4,834 | 10,488 | 8,957 | 8,589 | 3,916 |
| | C | 2,711 | 9,399 | 14,430 | 24,033 | 9,418 |
| | D | 1,656 | 4,337 | 6,563 | 22,282 | 28,540 |
| | E | 130 | 251 | 247 | 413 | 570 |

The second and third columns of Table 8 present the comparison of predictive power for the two evaluation indicators for all the publications of the dataset. Now we suppose that we want to identify (as in the VQR framework) the two best publications for each professor of the dataset, basing the choice on $C_{short}$ or on C-J: the third and fourth columns show the data on the grading when restricted to these subsets of publications (43,174 in all). $C_{short}$ always provides a greater share (28.9%) of correct grades than does C-J (25.4%), even though there is a notable increase in incidence of cases of overgrading (62.8% with $C_{short}$ and 58.9% with C-J). Such a trend was fully predictable given that these subsets are polarized towards products with high grading: for the products graded as "A" in particular, it is evident that the only possible error is that of overgrading, and not of undergrading.

*Table 8: Comparison between the VQR evaluations (A=excellent; B= good; C=adequate; D=poor E=very poor) by $C_{long}$ vs respectively by $C_{short}$ and C-J*

|  | All publications | | The best 2 publications | |
|---|---|---|---|---|
|  | $C_{short}$ | C-J | $C_{short}$ | C-J |
| Correct grading | 44.2% | 28.4% | 28.9% | 25.4% |
| Overgrading | 36.2% | 54.0% | 62.8% | 58.9% |
| Undergrading | 19.6% | 17.6% | 8.3% | 15.7% |

The selection based on $C_{short}$ thus seems capable of identifying a greater number of "best products" than does the selection based on C-J. This is also confirmed by the analyses of the subsets of the "best products" selected using the three indicators in consideration: Table 9 shows the amplitude of intersection between pairs of subsets, each constituted of the best 43,174 selections (2 per professor) on the basis of C-J, $C_{short}$ and $C_{long}$. We note that for all the GEVs, the intersection between the best products selected on the basis of $C_{short}$ and $C_{long}$ is systematically greater than the intersection of best products on the basis of C-J and $C_{long}$. Physics is the only GEV where the superiority of $C_{short}$ is less evident.

*Table 9: Intersection of the subset of best publications by $C_{long}$ with, respectively, the subset by C-J and $C_{short}$*

| GEV | C-J ∩ $C_{long}$ | $C_{short}$ ∩ $C_{long}$ |
|---|---|---|
| 1 - Mathematics and computer science | 66.7% | 70.4% |
| 2 - Physics | 45.3% | 45.4% |
| 3 - Chemistry | 51.7% | 55.6% |
| 4 - Earth sciences | 70.5% | 73.0% |
| 5 - Biology | 63.0% | 66.9% |
| 6 - Medicine | 61.2% | 65.6% |
| 7 - Agricultural and veterinary sciences | 63.5% | 68.6% |
| 8a - Civil engineering | 74.8% | 91.9% |
| 9 - Industrial and information engineering | 60.7% | 76.4% |
| Total | 60.2% | 63.5% |

## 5. Conclusions

Intuition is a key ingredient in research and development, however improvisation rarely pays. Confirmation arrives from the case of the impact indicator based on a weighted combination of citation and journal metric percentiles, conceived by ANVUR and applied to assess the performance of research institutions in the Italian national research assessment exercise, VQR 2011-2014. In this paper we have seen that in spite of the short citation window, the simple count of the citations still showed greater predictive power of the long-term impact of the publications than that of the ANVUR combined indicator. This results holds true in all scientific disciplines, for any year of publication beginning from a citation window of two years, and is still more confirmed with greater breadth of citation window. Further, the result holds equally true for the entire set of the researchers' publications or for the best two of each individual (as required for submission to the VQR 2011-2014 assessment exercise). The attempt to reinforce the limited predictive power of citations when the window is short, by combining information on the prestige of the publishing journal, has shown itself to be miserable failure. The weighted linear combinations of the citation and journal metric percentiles proposed by Anfossi et al. (2016) actually weakened the predictive power of the citation metrics alone.

The current authors (also Italian) are not surprised to witness the mistaken use of indicators to assess the country's research institutions. Instead, the surprise is that the indicator was proposed, published (and then adopted) without any sort of empirical demonstration or theoretical argument that would legitimate it, and in spite of the literature warning against the use of the journal impact factor for the evaluation of articles. While the co-authors of the proposal are neophytes in bibliometrics, we would at least expect that such senior scientists (Sergio Benedetto, ANVUR executive committee, responsible for the VQR; Giorgio Parisi, winner of the Max Plank medal for theoretical physics and other international awards) would insist on the scientific method as a constant of all their work.

It might appear that at this point a logical follow-up would be to inquire into the extent of the distorting effects from the C-J indicator on the performance scores and ranking of the Italian research institutions, particularly since a rising share of annual government financing depends on the results of the research assessment. However to any international readers we can signal that the VQR framework is already prey to so many and such limitations, as evidenced in the works cited in introduction (failure to consider all indexed products instead of just two; failure to consider product quality values in the continuous range; adoption of full counting of the submitted publications regardless of the number of co-authors and their position in the byline; exaggerated times and costs of execution) that an exercise to extricate and further quantify this particular damage would indeed be of scarce value. What we can say to such international readers, hoping they are not so deaf as ANVUR, is: "*Refrain from adopting the combination of citation and journal metrics to grade publications, as used in the Italian VQR 2011-2014.*"